\documentclass[aps,twocolumn,superscriptaddress,showpacs]{revtex4}%
\usepackage{amsmath}
\usepackage{amssymb}
\usepackage{graphicx}
\usepackage{color}%
\providecommand{\U}[1]{\protect\rule{.1in}{.1in}}

\begin{document}
\title{Perfect Transfer of enhanced entanglement and asymmetric steering in a cavity magnomechanical system}

\author{Yao-Tong Chen}
\affiliation{School of Physics and Center for Quantum Sciences, Northeast Normal University, Changchun 130117, China}
\author{Lei Du}
\affiliation{Beijing Computational Science Research Center, Beijing 100193, China}
\author{Yan Zhang}
\email{zhangy345@nenu.edu.cn}
\affiliation{School of Physics and Center for Quantum Sciences, Northeast Normal University, Changchun 130117, China}
\author{Jin-Hui Wu}
\affiliation{School of Physics and Center for Quantum Sciences, Northeast Normal University, Changchun 130117, China}

\date{\today }


\begin{abstract}
We propose a hybrid cavity magnomechanical system to realize and transfer the bipartite entanglements and EPR steerings between magnons, photons and phonons in the regime of the stability of the system.
As a $\mathcal{PT}$-symmetric-like structure exhibiting the natural magnetostrictive magnon-phonon interaction, our passive-active-cavity system can be explored to enhance
the robust distant quantum entanglement and generate the relatively obvious asymmetric (even directional) EPR steering that is useful for the task with the highly asymmetric trusts of the bidirectional local measurements between two entangled states.
It is of great interest that, based on such a tunable magnomechanical system, the perfect transfer between near and distant entanglements/steerings of different mode pairs is realized by adjusting the coupling parameters; especially, the perfect transfer scheme of steerings is first proposed here.
These transferring processes suggest indeed a novel method for the quantum information storage and manipulation. In addition, the entanglements and steerings can also be exchanged between different mode pairs by adjusting the detunings between different modes. This work may provide a potential platform for distant/asymmetric quantum modulation.
\end{abstract}
\maketitle

\section{Introduction}

The quantum phenomena have been widely studied in various optomechanical systems \cite{2014RMP,kippenberg2008cavity,2010OMIT,2011EIT}.
Recently, however, technologies engineered from the cavity magnomechanical system have flourished \cite{2012,2013,2014,2015magnetic,2015spin,2015coherent,2016cavity}, providing a fertile arena for the realization of the quantum coherence between magnons, cavity photons and phonons.
Based on collective excitations of the ferromagnetic spin system like the yttrium iron garnet (YIG), magnon can be freely used for realizing the strong coupling to cavities and superconducting qubits theoretically \cite{2013theo,2014theo,2016theo1,2016theo2} and experimentally \cite{2015exp,2017exp} owing to its low damping rate, rich magnonic nonlinearities \cite{2010YIG,2011YIG,2015YIG,2016YIG}.
Especially, it is viable to adjust the magnons' frequency via a bias magnetic field \cite{2016cavity}, and control the damping rate by a grounded loop antenna above the YIG sphere to be much weaker than that of cavities \cite{2020YIG}, which is beneficial to the precise measurement \cite{du2021controllable}.
As a kind of magnetic material with high spin density, YIG has a magnon mode and also hosts a magnomechanical vibrational phonon mode by the geometric deformation of the surface as an effective mechanical resonator.
Then, these two modes can couple with each other via the magnetostrictive interaction, which has almost been ignored in common dielectric or metallic materials \cite{2016cavity,2019lyx,2019sideband}.
Benefiting from this special feature, the interest in systems involving YIG has been raised and various coherent phenomena similar to those in optomechanical systems can be studied, such as magnomechanically induced transparency/absorption \cite{2016cavity,wang2018magnon}, the bistability in cavity magnon polaritons \cite{2018bistability}, magnon-induced nonreciprocity \cite{2019nonreciprocity1,2019nonreciprocity2}, high-order sidebands generation \cite{2019lyx,2019sideband}, and quantum entanglement \cite{2018PRL,2019entanglement1,2020PRL}.
Recently, hybrid systems involving magnons, have become a promising platform for implementing exceptional points \cite{2017ep,2019ep}, exceptional magnetic sensitivity of magnon polaritons \cite{2019sensitivity}, enhanced sideband responses \cite{2019lyx}, and magnetic chaos \cite{1994chaos,2019chaos1,2019chaos2}.
However, the cavity magnomechanics as a novel scheme requires more comprehensive and deeper understanding in physics.

As a vital quantum mechanical phenomenon, the entanglement has been realized in many sorts of systems at the mesoscopic level \cite{2001mesoscopic,2003mesoscopic,2013mesoscopic} or the microscopic level \cite{2009microscopic,2011microscopic,2017microscopic}.
It is regarded as a key resource required to operate a quantum computer and to communicate with security guaranteed by physics laws \cite{2012PRX}. The entanglement between remote/distant objects can also serve as quantum memories \cite{2018remote}.
In the optomechanics, the entanglement has been generated between phonons and photons \cite{2007entanglement}, photons \cite{2013entanglement}, and phonons \cite{2017microscopic}.
With atoms, the atom-cavity-mirror tripartite entanglement emerges \cite{2008tripartite,2015tripartite} and the entanglement transfer \cite{2004transfer,2017transfer} is observed.
Note that, in the cavity magnomechanical system, the magnon-photon-phonon tripartite entanglement has been proposed, where the magnon-phonon entanglement can transfer to other subsystems via the coupling \cite{2018PRL}.
Then, the magnon-magnon entanglement has been achieved and enhanced via a flux-driven Josephson parametric amplifier \cite{2019JPA}, Kerr nonlinear effect \cite{2019kerr}, and quantum correlated microwave fields \cite{2019squeezed}.

In recent years, one quantum inseparability called Einstein-Podolsky-Rosen (EPR) steering has become a hot topic \cite{steering1,steeringHe}.
In nature, there are three different types of nonlocal correlations: the entanglement, the Bell nonlocality that can be stated without any reference to quantum theory \cite{walborn2011revealing}, and the EPR steering as an intermediate quantum nonlocality between the first two \cite{2007steering}.
While describing the ability of one party to remotely affect another's state through local measurements, the EPR steering is a subset of the entanglement and stronger than the entanglement \cite{2015steering,2019steering}.
Essentially distinct from the entanglement, it can feature a unique asymmetry between two observers under proper conditions \cite{steering2,steering3,steeringarxiv}, which has been experimentally demonstrated \cite{saunders2010experimental}.
Moreover, for experimental implementations, the quantum-refereed steering, only requiring one joint measurement, might be considerably easier to test than the entanglement witnessing requiring two joint measurements \cite{cavalcanti2013entanglement}.
Thus, the observation of steering has been used for detecting entanglements in some systems, such as Bose-Einstein condensates \cite{he2011einstein} and atomic ensembles \cite{he2013towards}.
Being important for explaining basic characteristics of the quantum mechanics, the asymmetric steering can also be used to complete the task involving highly asymmetric quantum modulation and realize many quantum information protocols with super security.
Recently, the asymmetric steering has shown several significant applications, e.g., cryptography \cite{branciard2012one}, randomness generation \cite{law2014quantum}, channel discrimination \cite{piani2015necessary} and tele-amplification \cite{he2015secure}.
However, compared with the entanglement, the generation of asymmetric steering requires relatively strict conditions, such as a stronger two-mode quantum correlation and asymmetric mean quantum numbers of two modes which determines the steering direction \cite{2019steering}.

On the other hand, the non-Hermitian parity-time ($\mathcal{PT}$) symmetry has been exploited to explore and enhance some interesting quantum phenomena, and the optical $\mathcal{PT}$-symmetric system can be realized using microcavity or optomechanical systems \cite{li2015proposal,2019delayed}.
One important effect in $\mathcal{PT}$-symmetric cavity system is the photonic localization around the $\mathcal{PT}$-phase-transition point, which can induce the dynamical accumulations of optical energy in the two-supermode-based cavities and then be used to enhance the photon blockade \cite{li2015proposal} and generate unidirectional phonon transport \cite{zhang2015giant}.
Note that the optical $\mathcal{PT}$-symmetry requires the exact balance between the loss and gain.
However, the enough strong gain may break the stationary response of the system, and the balance condition may also be too strict for the realistic implementation especially when tiny disturbances are inevitable.
The $\mathcal{PT}$-symmetric-like system not requiring the strict balance can still follow the predictions of the true $\mathcal{PT}$ one in many cases, and thus attract considerable attentions \cite{2017lyx,2019lyx}.
Identical with the $\mathcal{PT}$-symmetric system, the real/imaginary parts of the Hamiltonian eigenvalues of the $\mathcal{PT}$-like one can spontaneously coalesce or separate, i.e., there is a transition point between the unbroken and broken $\mathcal{PT}$-like phases.
Moreover, such a system can also show the photonic localization effect for some useful applications, such as the enhancement of the optical linear \cite{2017lyx} and nonlinear \cite{gePTlike} transmissions and the significant light group delay \cite{2019lyx}.
Therefore, we try to use the $\mathcal{PT}$-like system for enhancing the quantum entanglement and satisfying the requirements of the creation of the EPR steering.

With a four-mode magnomechanical system of $\mathcal{PT}$-symmetric-like scheme by adding an auxiliary active cavity, we would realize robust enhanced bipartite entanglements and asymmetric EPR steerings between magnon, phonon and photon modes. We just add a cavity to the basic magnomechanical system of a cavity coupling a YIG sphere, which seems a little complex but is necessary for our aims. Here we focus on the underlying physical understanding of the creation of entanglements/steerings in the cavity magnomechanics, and consequently find out the mechanism that how the incoherent gain process affects the entanglement arising from the coherent nonlinear coupling.
Furthermore, such a well-designed double-cavity system can be used to entangle/steer distant subsystems, and then the perfect transfer from the initial near entanglement/steering to the distant one can be realized due to the tunability of the system.
For the quantum-communication network \cite{genes2008robust}, our system, as a basic component, could be used to transfer quantum information/modulation involving the entanglement/steerability further via adding more sites for being extended to a communicating chain with a relatively high efficiency.
To the best of our knowledge, our work is the first one that focuses on an efficient scheme for realizing the perfect transfer of quantum steering.

\section{The Model and method}

\begin{figure}[ptb]
\centering
\includegraphics[width=8.6 cm]{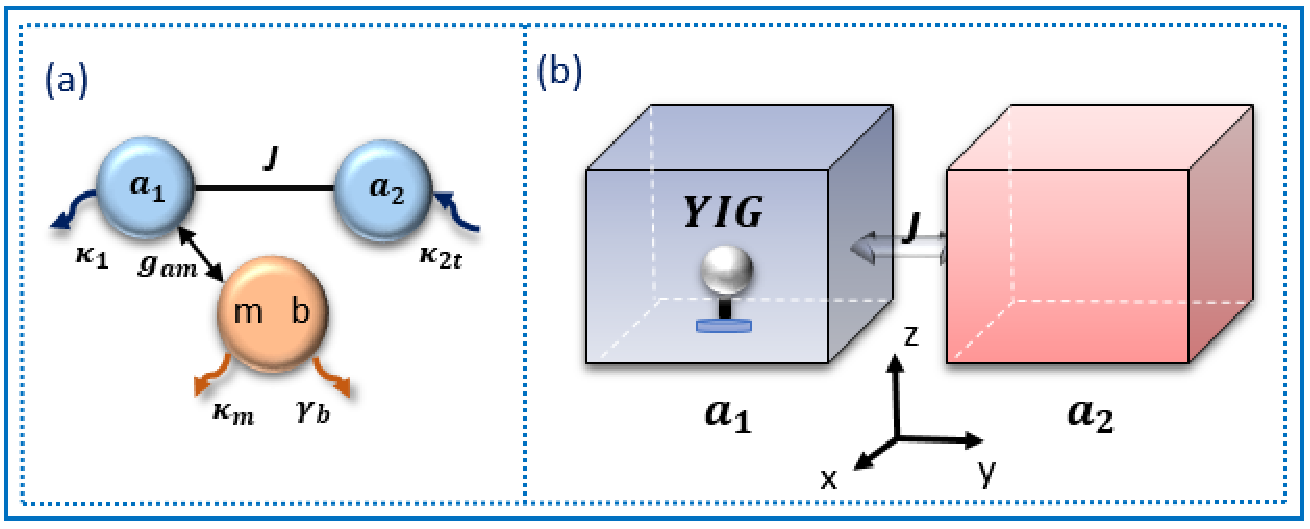}
\caption{(Color online) (a) Schematic diagram of a passive-active double-cavity magnomechanical system, a YIG sphere is in the passive cavity $a_{1}$ coupled with the active cavity $a_{2}$ directly. (b) Implementation of the magnomechanical system with the photon-tunneling strength $J$ between two three-dimensional cavities.}
\label{fig1}
\end{figure}

As schematically shown in Fig.~\ref{fig1}, we consider a cavity magnetomechanical system including two three-dimensional microwave cavities coupled with each other via optical tunneling mutually, of which coupling strength $J$ can be tuned by the adjustment of the distance between two cavities. Cavity $a_{1}$ ($a_{2}$) is passive (active), and then such a $\mathcal{PT}$-symmetric-like structure may further affect the properties of this cavity magnetomechanical system.
A YIG sphere of volume $\mathcal{V}$, whose size is smaller than the wavelength of the cavity field, is put into cavity $a_{1}$ and near the maximum of the magnetic field of $a_{1}$ along the $x$-axis.
Via the magnetostrictive effect, the YIG is driven by a strong driving magnetic field of amplitude (frequency) $B_{0}$ ($\omega_{0}$) along the $y$-axis for generating a magnomechanical vibrational phonon mode, which is nonlinearly coupled with the magnon mode \cite{2016cavity,2018PRL}.
The photon mode of the cavity and the magnon mode of the YIG are coupled with each other via magnetic-dipole interaction induced by a uniform bias magnetic field of intensity $H$ along the $z$-axis, where the coupling strength is tunable by adjusting the intensity of the bias magnetic field or the position of YIG.

The Hamiltonian of such a system under the rotating-wave approximaton can be written as $(\hbar=1)$
\begin{eqnarray}
&\mathcal{H}=\omega_{1}a_{1}^{\dagger}a_{1}+\omega_{2}a_{2}^{\dagger}a_{2}+\omega_{m}m^{\dagger}m+\mathcal{K}m^{\dagger}mm^{\dagger}m\nonumber\\
&+\frac{\omega_{b}}{2}(q^{2}+p^{2})\nonumber+g_{mb}m^{\dagger}mq+g_{ma}(a_{1}m^{\dagger}+a_{1}^{\dagger}m)\nonumber\\
&\quad\,\,\,+J(a_{1}a_{2}^{\dagger}+a_{1}^{\dagger}a_{2})+i\varepsilon_d(m^{\dagger}e^{-i\omega_{0}t}-H.c.),
\label{eq1}
\end{eqnarray}
where $\omega_{i}$ $(i=1,2)$, $\omega_{m}$, and $\omega_{b}$ are resonant frequencies of the cavity photon modes, the magnon mode, and the mechanical phonon mode, respectively.
$a_{i}/a_{i}^{\dagger}$, $m/m^\dagger$ and $b/b^{\dagger}$ are, respectively, the annihilation/creation operators of the cavity, magnon, phonon modes, and satisfy the commuting relation $[o, o^{\dagger}]=1$ $(o=a_{i}, b, m)$.
$q=(b+b^{\dagger})/\sqrt{2}$ and $p=i(b^{\dagger}-b)/\sqrt{2}$ are dimensionless position and momentum quadratures of the phonon mode.
And the frequency $\omega_{m}$ of the magnon mode is defined as $\omega_{m}=H \gamma_{g}$ with the bias magnetic filed  and the gyromagnetic ratio of the YIG sphere $\gamma_{g}/2\pi=28$ GHz/T \cite{2018PRL}.
The well-accepted Young's modulus of the YIG is about $Y=2\times10^{11}$ Pa \cite{gibbons1958acoustical}, the Poisson ratio is about $v=0.29$, the spin density $\rho=4.22\times10^{27}$ m$^{-3}$, the spin number $S=\frac{5}{2}$ \cite{2016cavity}, and the total number of spins $N=\rho \mathcal{V}$.
In the interaction terms of Hamiltonian, $g_{ma}$ ($g_{mb}$) is the coupling strength of the single-magnon-photon (single-magnon-phonon) interaction.
Here, $g_{ma}$, depending on the position of YIG \cite{str1,str2}, is assumed in the strong-coupling regime, which may promote the perfect transfer of the entanglement and steering, as $g_{ma}>\kappa_{1}, \kappa_{m}$, with $\kappa_{1}$ ($\kappa_{m}$) being the decay rate of the photon (magnon) mode;
whereas, $g_{mb}$ is much weaker and related to the spin density and spin number \cite{2016cavity}.
The Rabi frequency $\varepsilon_d=\frac{\sqrt{5}}{4}\gamma_{g}\sqrt{N}B_{0}$ \cite{2018PRL} determines the coupling strength between the driving field $B_{0}$ and the magnon.
The term $\mathcal{K}m^{\dagger}mm^{\dagger}m$ with the Kerr nonlinear coefficient $\mathcal{K}$ describes the Kerr nonlinear effect that can be neglected under the condition of $\mathcal{K}|\langle m \rangle|^{3}\ll\varepsilon_{d}$ \cite{2018bistability}.
Specifically, for this approximation, choosing a proper volume $\mathcal{V}$ is important in the experimental implementation. The reason is that $\mathcal{K}$ is inversely proportional to $\mathcal{V}$ \cite{2018PRL} but the magnomechanical coupling strength is obviously weakened with the increasing of $\mathcal{V}$ in terms of the experimental results \cite{2016cavity}.
Here we choose a proper sphere of diameter $D= 250$ $\mu$m so as to neglect the Kerr nonlinearity as used in the experiment \cite{2016magnon}.
In addition, the other nonlinear effect considered here, i.e., the natural magnetostrictive magnon-phonon interaction within the YIG sphere, plays a key role in the generation of the entanglement and steering, which will be discussed in the next section.

In the rotating frame with respect to the driving filed frequency $\omega_{0}$, the quantum Langevin equations for the operators in the system with the relevant dissipations and noises can be obtained as
\begin{equation}
\begin{split}
\dot{a_{1}}&=-(i\Delta_{1}+\kappa_{1})a_{1}-ig_{ma}m-iJa_{2}+\sqrt{2\kappa_{1}}a_{1,in}\\
\dot{a_{2}}&=-(i\Delta_{2}+\kappa_{2t})a_{2}-iJa_{1}+\sqrt{2\kappa_{2}}a_{2,in}+\sqrt{2g}a_{2,in}^{(g)}\\
\dot{m}&=-(i\Delta_{m}+\kappa_{m})m-ig_{ma}a_{1}-ig_{mb}mq+\varepsilon_d\\
&\,\,\,\,\,\,\,+\sqrt{2\kappa_{m}}m_{in}\\
\dot{q}&=\omega_{b}p\\
\dot{p}&=-\omega_{b}q-\gamma_{b}p-g_{mb}m^{\dagger}m+\xi,
\label{eq2}
\end{split}
\end{equation}
with $\Delta_{i}=\omega_{i}-\omega_{0}$ (i=1,2) and $\Delta_{m}'=\omega_{m}-\omega_{0}$ being the relevant detunings. $\kappa_{2t}=\kappa_{2}-g$ is the effective damping rate of cavity $a_{2}$, where $g$ is the real gain into cavity $a_{2}$.
Via setting $\eta \equiv \kappa_{2t}/\kappa_{1}$, $a_{2}$ is an active (passive) cavity with $\eta<0$ ($\eta>0$). $a_{i,in}$, $m_{in}$ and $\xi$ are the noise operators associated with the photon, magnon and phonon modes with zero mean values, and are characterized by nonzero time-domain correlation functions \cite{book} as, respectively, $\langle k_{in}(t)k_{in}^{\dagger}(t')\rangle=[N_{k}(\omega_{k})+1]\delta(t-t')$, $\langle k_{in}^{\dagger}(t)k_{in}(t')\rangle=N_{k}(\omega_{k})\delta(t-t')$ with $k=a_{1}, a_{2}, m$;
the noise operators associated with the gain in cavity $a_{2}$ are \cite{2018jiangcheng} $\langle a_{2,in}^{(g)}(t)a_{2,in}^{(g)\dagger}(t')\rangle=N_{a_{2}}(\omega_{a_{2}})\delta(t-t')$, $\langle a_{2,in}^{(g)\dagger} (t)a_{2,in}^{(g)}(t')\rangle=[N_{a_{2}}(\omega_{a_{2}}+1)]\delta(t-t')$;
under a large mechanical quality factor $Q_{m}=\omega_{b}/\gamma\gg1$ \cite{1981}, the Langevin force operator $\xi$ is simplified as a $\delta$-correlated function with the Markovian approximation, i.e.,
$\langle\xi(t)\xi(t')+\xi(t')\xi(t)\rangle/2\simeq\gamma_{b}[2N_{b}\omega_{b}+1]\delta(t-t')$.
Here $N_{k}(\omega_{k})=[exp(\frac{\hbar\omega_{k}}{k_{B}T})-1]^{-1}$ and $N_{b}(\omega_{b})=[exp(\frac{\hbar\omega_{b}}{k_{B}T})-1]^{-1}$ are the mean thermal excitation numbers in the environmental temperature $T$, where $k_{B}$ is the Boltzmann constant.
The bipartite steady-state entanglement between modes $M_{1,2}$ can be quantified by the logarithmic negativity $E_{M_{1}M_{2}}$ arising from the covariance matrix (CM) which is obtained by solving Eq.~(\ref{eq9}), and the EPR steering can be quantified by $\zeta^{M_{1}\rightarrow M_{2}}$ or $\zeta^{M_{2}\rightarrow M_{1}}$ with Eq.~(\ref{eq10}). Calculation methods of these quantities are detailed in Appendix.

Note that, before quantifying the entanglement in the steady state, the analysis of the stable parameter regimes of the system ought to be done.
The stability refers to the existence of asymptotic steady state of the system, i.e., the system can be retained at a steady state for a long evolution time. And then, in the parameter regime of the stability, the stationary response of the system can be observed.
When being unstable, the system can evolve toward another state and may even show the oscillation behavior between different states.
For steady-state entanglements and steerings, in our system, the steady-state covariance matrix $V$ in Eq.~(\ref{eq8}) is required to be asymptotically stable.
For finding the stable regime, based on the Routh-Hurwitz criterion \cite{dejesus1987routh}, we need to clarify the real part sign of the eigenvalues of the Jacobian matrix $A$ as shown in Eq.~(\ref{eq6}) in Appendix.
Only if the real parts of all eigenvalues are negative, the system is stable \cite{dejesus1987routh, dumeige2011stability}.
Finally, with such parameters of stable region, the steady-state entanglement can be generated.

\section{Results and discussions}

\begin{figure}[ptb]
\centering
\includegraphics[width=8.6 cm]{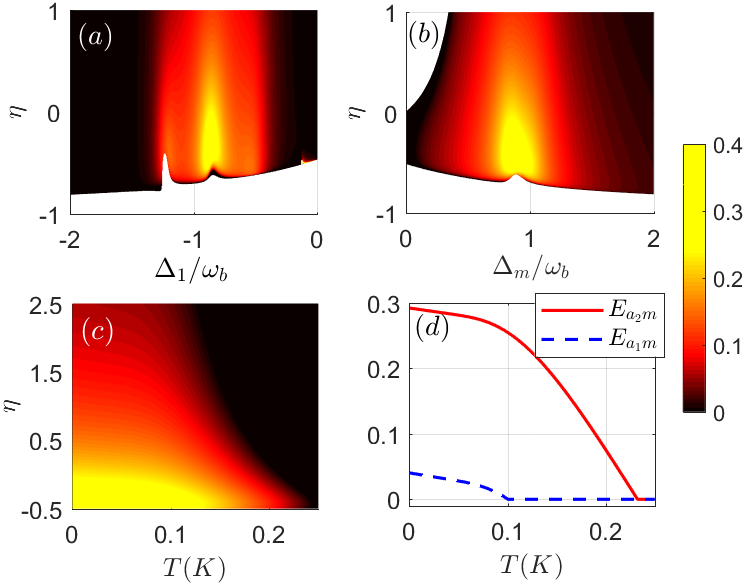}
\caption{(Color online) Logarithmic negativity of the entanglement (a) $E_{a_{2}m}$ versus $\Delta_{1}$ and $\eta$ with $\Delta_{m}=0.9\omega_{b}$; (b) $E_{a_{2}m}$ versus $\Delta_{m}$ and $\eta$ with $\Delta_{1}=-0.91\omega_{b}$; (c) $E_{a_{2}m}$ versus $T$ and $\eta$ with $\Delta_{1}=-0.91\omega_{b}$ and $\Delta_{m}=0.89\omega_{b}$; (d) $E_{a_{1}m}$ and $E_{a_{2}m}$ versus $T$ with $\Delta_{1}=-0.91\omega_{b}$, $\Delta_{m}=0.89\omega_{b}$ and $\eta=-0.5$. The environment temperature $T=15$ mK in (a) and (b). Here $\Delta_{1}=\Delta_{2}$, $J=2\kappa_{1}$, and $\eta \equiv \frac{\kappa_{2t}}{\kappa_{1}}$ is the dimensionless ratio of the loss (gain) of two cavities.}
\label{fig2}
\end{figure}

In our scheme, it is worth to note that, via the addition of the auxiliary cavity $a_{2}$, the distant bipartite photon-magnon entanglement $E_{a_{2}m}$ can be generated and then be modulated to be more salient than the near one $E_{a_{1}m}$.
The experimentally reachable parameters \cite{2016cavity,2018PRL} in our system are: $\omega_{b}/2\pi=10$ MHz, $\omega_{1}/2\pi=\omega_{2}/2\pi=10$ GHz, $\kappa_{1}/2\pi=1$ MHz, $\kappa_{m}/2\pi=0.56$ MHz, $\gamma_{b}/2\pi=10^{2}$ Hz, $g_{ma}/2\pi=3.2$ MHz, $g_{mb}/2\pi=0.2$ MHz, and the driving magnetic field $B_{0}=3.9\times10^{-5}$ T corresponding to the driving power $P=8.9$ mW with the relation $P=\frac{B_{0}^2\pi(D/2)^2 c}{2\mu _0}$ .

Here the magnon mode $m$ is driven to the near-resonance on the blue (anti-Stokes) sideband $\omega_{0}+\omega_{b}$; then the cavities $a_{1}$ and $a_{2}$ are both near-resonant on the red (Stokes) sideband $\omega_{0}-\omega_{b}$.
The nonlinear magnomechanical coupling between magnon and phonon modes plays a key role in the generation of the entanglement/steering.
With the fluctuation operators as Eq. (A2) in Appendix, the magnon-phonon interaction term of Hamiltonian, i.e., the sixth term in Eq. (1), can be rewritten and decomposed into two terms as $G_{mb}(\delta m^{\dagger}\delta b+\delta m\delta b^{\dagger})+G_{mb}(\delta m\delta b+\delta m^{\dagger}\delta b^{\dagger})$ with the effective magnomechanical coupling coefficient $G_{mb}=i\sqrt{2}g_{mb}\langle m\rangle$ as in Eq. (A4) in Appendix.
The first beam-splitter interaction term can be used to cool phonons and to generate a state swap between the magnon and phonon modes; the second two-mode-squeezing interaction term describes a magnomechancal analog to the nonlinear optical down-conversion process and then create the initial entanglement from coherent input states, i.e., phonons and magnons in pairs \cite{2014RMP,2013mesoscopic}. Then the initial magnon-phonon entanglement can transfer to other subsystems via coupling, i.e., magnon-photon, photon-phonon and photon-photon entanglements. Compared with such a magnomechanical system, the generation in a linear-coupling cavity/magnon system may need the injection of external two-mode squeezed fields for generating entanglements \cite{2019squeezed} or other conditions.

In Figs.~\ref{fig2}(a) and (b), we plot the distant entanglement $E_{a_{2}m}$ in the parameter regime of the stable system. With the decreasing of $\eta$, the entanglement $E_{a_{2}m}$ gradually increases, especially when $\eta \equiv \kappa_{2t}/\kappa_{1}<0$ (introducing gain to $a_{2}$).
This means the entanglement arising from the \textit{coherent} nonlinear coupling can be enhanced by the \textit{incoherent} gain process in the $\mathcal{PT}$-symmetric-like structure.
The reason is that the increasing number of photons in the active cavity $a_{2}$ corresponding to the increasing gain may induce the enhancement of the entanglement $E_{a_{2}m}$ even if this entanglement is distant.
Specifically, the gain introduced increases the mean number of photons $|\langle a_{2}\rangle|^{2}$ in the active cavity $a_{2}$, and then the energy can transfer from $a_{2}$ to the passive cavity $a_{1}$ and the magnon $m$ via the coupling.
The dependence of the mean number of magnons $|\langle m\rangle|^{2}$ on the gain can be observed analytically as Eq.~(\ref{eq4}) in Appendix.
According to Eq.~(\ref{eq4}), $|\langle m\rangle|$ becomes large gradually with the increasing gain, i.e., the decreasing of $\kappa_{2t}$. And then the effective magnomechanical coupling strength $|G_{mb}|=\sqrt{2}g_{mb}|\langle m\rangle|$ also becomes large in the parameter range.
This implies that, in the effective nonlinear magnetostrictive coupling between magnons and phonons, the incoherent gain can strengthen two-mode-squeezing interaction so that the initial entanglements can be enhanced obviously, which can be transferred to other subsystems.
These can also be shown in relevant equations, i.e., Eqs.~(\ref{eq7})-(\ref{eq9}). The mangon-phonon entanglement $E_{mb}$ arises from relevant entries of the real covariance matrix $V$ including $|G_{mb}|$.
However, for maintaining the system stability, the gain cannot be too large to satisfy the ideal gain-loss balance, thus such a system is in the regime of $\mathcal{PT}$-symmetric-like scheme.
Meanwhile, when the gain goes up, the near entanglement $E_{a_{1}m}$ can also increase because the photons also transfer from $a_{2}$ to $a_{1}$ via the tunnelling.
Moreover, the maximum of $E_{a_{2}m}$ can reach about $0.45$ and thus is more than twice as large as that of $E_{a_{1}m}$.
Thus, the $\mathcal{PT}$-symmetric-like scheme of such a system can raise the possibility to obtain more magnon-photon pairs, and it may enhance both the distant and near entanglements.

\begin{figure}[ptb]
\centering
\includegraphics[width=8.6 cm]{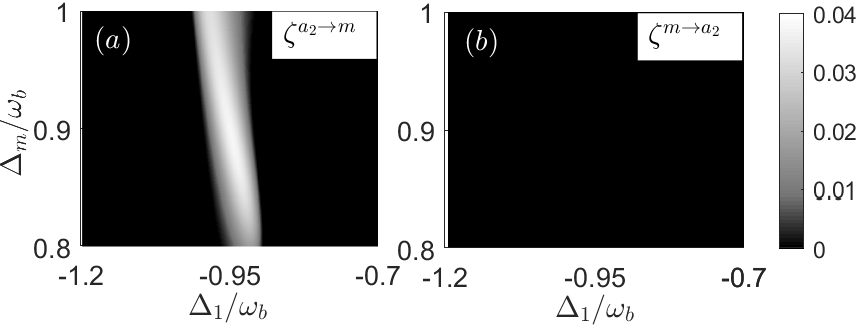}
\caption{(Color online) Quantum steering (a) $\zeta^{a_{2}\rightarrow m}$ and (b) $\zeta^{m\rightarrow a_{2}}$ versus $\Delta_{1}$ and $\Delta_{m}$ with $\eta=-0.5$. The other parameters are the same as those in Fig.~\ref{fig2}.}
\label{fig3}
\end{figure}

Figure~\ref{fig2}(c) shows the robustness of the distant entanglement $E_{a_{2}m}$ against the environment temperature $T$.
As $\eta$ decreases, $E_{a_{2}m}$ becomes more obvious, especially when $\eta<0$.
Meanwhile, note that the critical temperature, below which the entanglement $E_{a_{2}m}$ appears ($E_{a_{2}m}>0$), increases gradually; when $\eta=-0.5$, the entanglement can be observed until $T\simeq0.25K$.
The $\mathcal{PT}$-symmetric-like passive-active cavity system exhibits a superior performance than the passive-passive one ($\eta>0$), of which the critical temperature is about $0.1K$.
In addition, the robustness of the near entanglement can also be strengthened in such a system (not shown here).
However, note that the critical temperature of the distant entanglement $E_{a_{2}m}$ is much higher than that of the near entanglement $E_{a_{1}m}$ in Fig.~\ref{fig2}(d) with $\eta=-0.5$.
At a proper temperature, the distant entanglement with suppressed near one would be an advantage.
As for the low temperature environment, we can use a dilution fridge to accommodate the system \cite{2020PRL}.
However, in such a $\mathcal{PT}$-symmetric-like system, the gain introduced also enhances the robustness against the temperature, i.e., the entanglement of a relatively large degree can be observed in a wider temperature range.
Thus, this $\mathcal{PT}$-symmetric-like scheme reduces the need for harsh temperature of generating entanglement, which may be conducive to the actual experiment.

In such a system, there may be asymmetric two-way (even directional) EPR steering between two modes of the subsystem due to the different decay/gain in the $\mathcal{PT}$-symmetric-like scheme and the different coupling sidebands in the anti-Stokes/Stokes processes.
As shown in Fig.~\ref{fig3}, we obtain the directional distant steering $a_{2}\rightarrow m$ from photon mode $a_{2}$ to magnon mode $m$, described by $\zeta^{a_{2}\rightarrow m}\neq0$ and $\zeta^{m\rightarrow a_{2}}=0$, i.e., $a_{2}$ can steer $m$ when the steerability disappears in the opposite direction.
Whereas, in the passive-passive system, the steerability is very tiny (not shown here).
Generally, the addition of the losses or thermal noises can lead to the one-way steering \cite{steering2,steering4}.
However, in our model, the gain can bring the obvious steering, which implies the feasibility for adjusting (steering) the state of the YIG sphere by the auxiliary cavity $a_{2}$.
In addition, such a directional steering is obtained when the magnon and photon modes are driven on different sidebands for the appearance of the obvious distant entanglement.
Generally speaking, the creation of the asymmetric steering requires an enough strong two-mode quantum correlation and the asymmetric mean numbers of two modes, both of which can be well satisfied in our $\mathcal{PT}$-symmetric-like system.
The reason is that as the gain in the auxiliary cavity $a_{2}$ strengthen the two-mode correlation via the two-mode-squeezing interaction, such a non-Hermitian $\mathcal{PT}$-symmetric-like system composed of asymmetric spatial distribution of the loss and gain can show obvious difference between the mean quantum numbers of two modes, which is caused by the photonic localization effect \cite{zhang2015giant}.

\begin{figure}[ptb]
\centering
\includegraphics[width=8.6 cm]{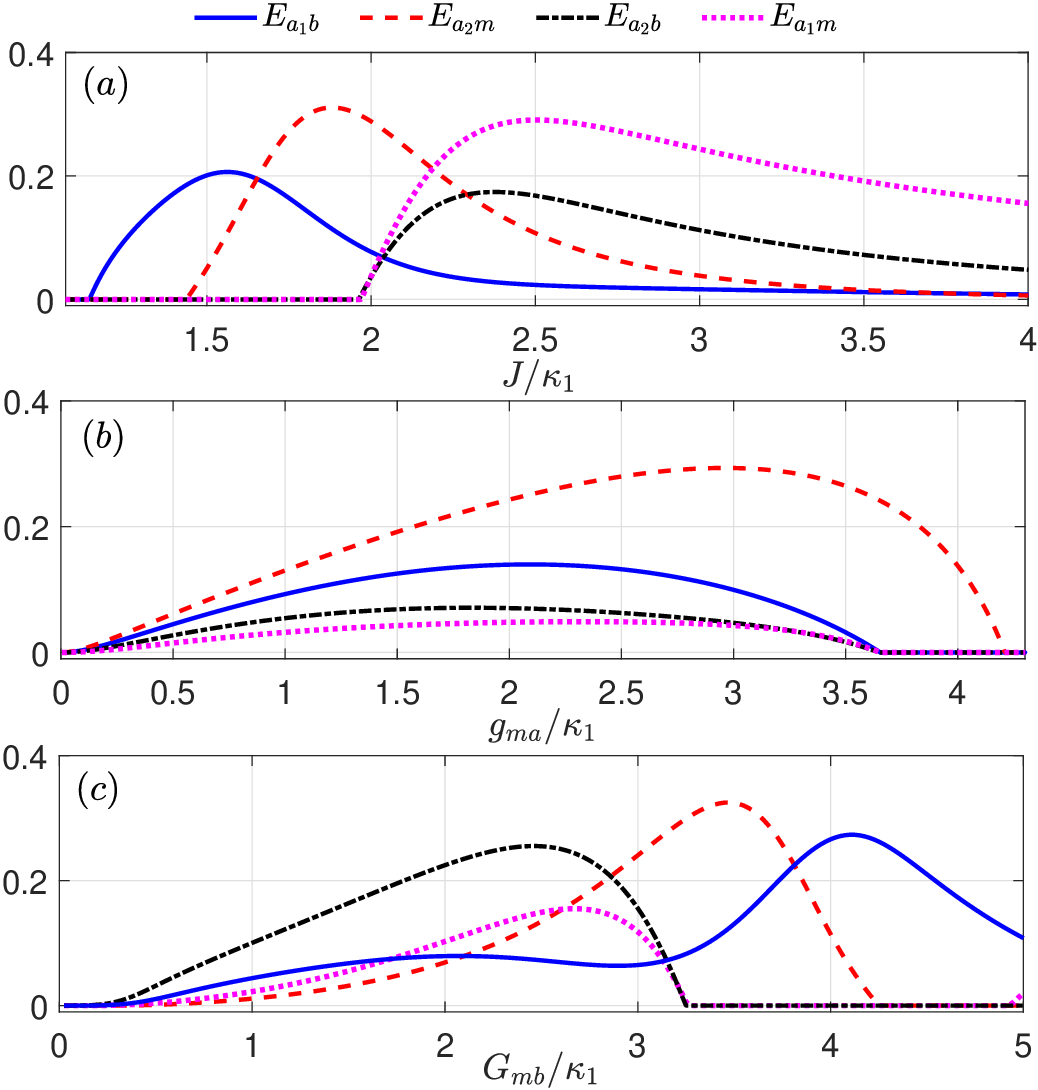}
\caption{(Color online)  Logarithmic negativity of the entanglement $E_{a_{1}b}$, $E_{a_{2}b}$, $E_{a_{1}m}$ and $E_{a_{2}m}$ versus (a) photon-photon coupling strength $J$; (b) magnon-photon coupling strength $g_{ma}$; (c) effective magnon-phonon coupling strength $G_{mb}$ with $\Delta_{1}=-0.91\omega_{b}$, $\Delta_{m}=0.89\omega_{b}$, and $\eta=-0.5$. The other parameters are the same as those in Fig.~\ref{fig2}.}
\label{fig4}
\end{figure}

The perfect transfer between entanglements of various modes (distant and near entanglements) is very important for the quantum information processing and transmission \cite{2018PRL,2004transfer,2017transfer}.
Figure~\ref{fig4} shows the magnon-photon entanglements $E_{a_{2}m}$, $E_{a_{1}m}$, and the magnon-phonon entanglement $E_{a_{2}b}$, $E_{a_{1}b}$ in the subsystems with the varying cavity-cavity coupling strength $J$, magnon-cavity coupling one $g_{ma}$ and mangon-phonon coupling one $G_{mb}$.
The photon-photon entanglement $E_{a_{1}a_{2}}$ constantly disappears with the parameters used (not shown here).
In Fig.~\ref{fig4}(a), it is clear that the near photon-phonon entanglement $E_{a_{1}b}$ peaks at $J=1.6\kappa_{1}$; then, with the increasing of $J$, while $E_{a_{1}b}$ decreases, the distant mangon-photon entanglement $E_{a_{2}m}$ appears, increases, exceeds $E_{a_{1}b}$ at $J=1.7\kappa_{1}$ and peaks at $J=1.9\kappa_{1}$.
That process can be regarded as the transfer between $E_{a_{2}m}$ and $E_{a_{1}b}$, i.e., between distant and near entanglements of different modes.
When both $E_{a_{1}b}$ and $E_{a_{2}m}$ decrease, both the near entanglement $E_{a_{1}m}$ and the distant one $E_{a_{2}b}$ appear at $J=1.97\kappa_{1}$; then, with $J$ becomes big enough, there are still $E_{a_{2}b}$ and $E_{a_{1}m}$ when both $E_{a_{1}b}$ and $E_{a_{2}m}$ nearly disappear. That process implies the perfect entanglement transfer, i.e., one entanglement tends to die while the other one appears, by adjusting coupling strength.
Except $J$, the coupling $g_{ma}$ can also adjust the entanglement.
As shown in Fig.~\ref{fig4}(b), four entanglements all appear from $g_{ma}=0$ ; but $E_{a_{1}b}$, $E_{a_{2}b}$ and $E_{a_{1}m}$ die at $g_{ma}=3.6\kappa_{1}$, and only distant entanglement $E_{a_{2}m}$ exists in the range  $g_{ma}/\kappa_{1}\in [3.6, 4.2]$.
When analyzing the effect of the effective magnon-phonon coupling $G_{mb}$ in Fig.~\ref{fig4}(c), we can also realize the transfer between different entanglements.
Those refer to that the stationary entanglement can be transferred from the subsystems at an obvious degree due to the transfer of photons, phonons and magnons among subsystems.
Note in particular that, when $G_{mb}=0$, there exists no entanglement due to the vanishing of the two-mode-squeezing interaction.
In those processes of controllable entanglements, with the gain into cavity $a_{2}$, it can be maintained that photon $a_{2}$ and magnon $m$ can show the biggest entanglement among all subsystems in the presence of abundant photons.

\begin{figure}[ptb]
\centering
\includegraphics[width=8.6 cm]{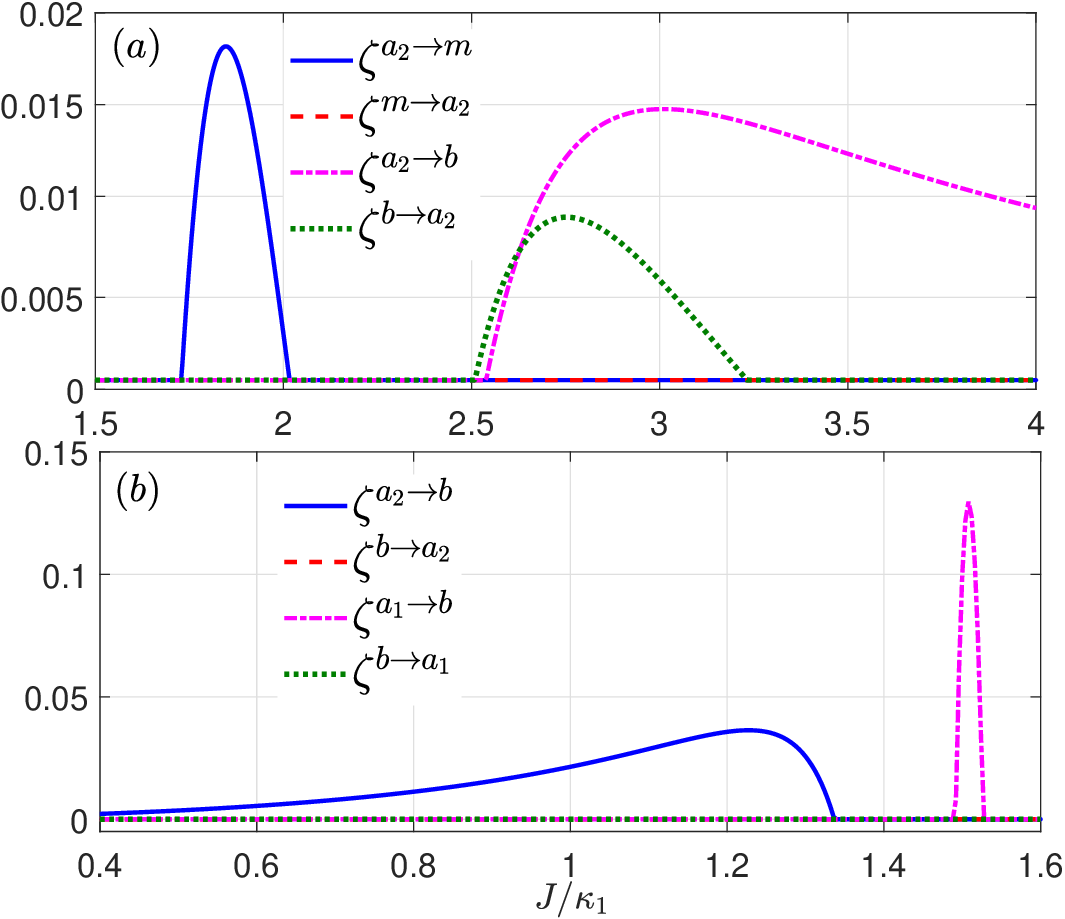}
\caption{(Color online) Quantum steerings between photon $a_{1}$, $a_{2}$, mangon $m$ and phonon $b$ versus $J$ with (a) $\Delta_{1}=-0.91\omega_{b}$, $\Delta_{m}=0.89\omega_{b}$, and $\Delta_{1}=\Delta_{2}$; (b) $\Delta_{1}=0.06\omega_{b}$, $\Delta_{m}=0.375\omega_{b}$, and $\Delta_{1}=-\Delta_{2}$. Here $\eta=-0.5$ and the other parameters are the same as those in Fig.~\ref{fig2}.}
\label{fig5}
\end{figure}

Moreover, it is worth noting that the perfect transfer of the asymmetric quantum EPR steering can also be realized in such a tunable hybrid system.
In Fig.~\ref{fig5}, we plot the steerings as functions of the cavity-cavity coupling strength $J$.
With $\eta=-0.5$, the maximum of the steering $\zeta^{a_{2}\rightarrow m}$ appears in the vicinity of $J=1.8\kappa_{1}$ corresponding to the entanglement $E_{a_{2}m}$ as shown in Fig.~\ref{fig2}(a), however, the steering of the opposite direction $\zeta^{m\rightarrow a_{2}}=0$.
With the increasing of $J$, the directional steering $a_{2}\rightarrow m$ dies and the asymmetric two-way steering between $a_{2}$ and $b$ appears from $J=2.5\kappa_{1}$.
And when $J/\kappa_{1}\in[2.5,3.2]$, the asymmetric two-way steering between $a_{2}$ and $b$ is observed with the suppressed steering between $a_{2}$ and $m$, which means the perfect transfer of steerings.
That process implies that we realize the transfer of directional steering between the photon-magnon pair and photon-phonon pair.
Figure~\ref{fig5}(b) shows the transfer from directional steering $a_{2} \rightarrow b$ to diectional steering $a_{1} \rightarrow b$, i.e., the perfect transfer between distant and near directional steerings.

\begin{figure}[ptb]
\centering
\includegraphics[width=8.6 cm]{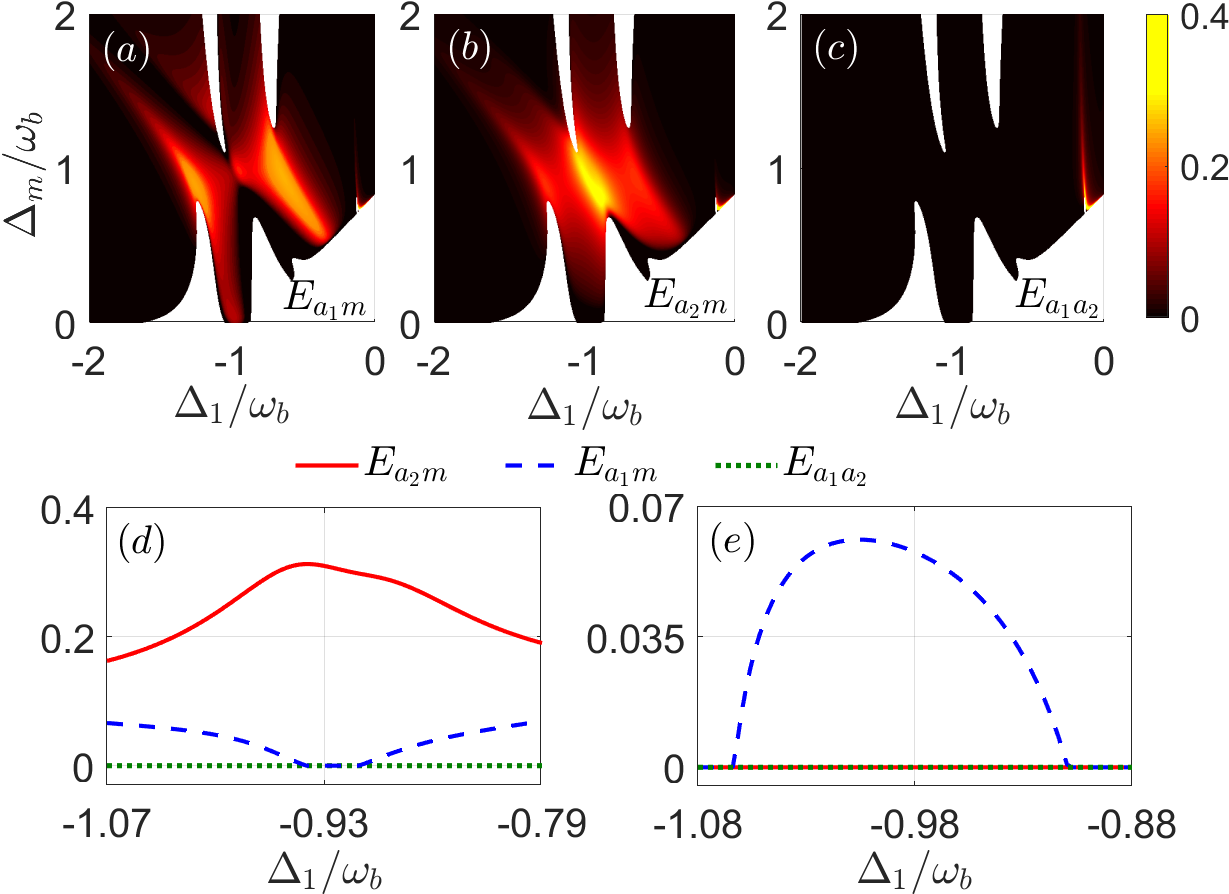}
\caption{(Color online) Logarithmic negativity of the entanglement (a) $E_{a_{1}m}$, (b) $E_{a_{2}m}$, and (c) $E_{a_{1}a_{2}}$ versus $\Delta_{1}$ and $\Delta_{m}$.  $E_{a_{1}m}$, $E_{a_{2}m}$ and $E_{a_{1}a_{2}}$ versus $\Delta_{1}$ with (d) $\Delta_{m}=0.87\omega_{b}$ and (e) $\Delta_{m}=0.06\omega_{b}$. Here $\eta=-0.5$ and the other parameters are the same as those in Fig.~\ref{fig2}.}
\label{fig6}
\end{figure}

To further understand the modulation of the near/distant entanglements/steerings, we plot the entanglements/steerings as functions of detunings. In Figs.~\ref{fig6}(a) and~\ref{fig6}(c), we plot the entanglements against the detunings $\Delta_{1}$ and $\Delta_{m}$ with $\eta=-0.5$ in the stable region of the system.
The two photon-magnon entanglements $E_{a_{1}m}$ and $E_{a_{2}m}$ are obvious in the vicinity of $\Delta_{1}\simeq -\omega_{b}$ and $\Delta_{m}\simeq \omega_{b}$, while photon-photon entanglement $E_{a_{1}a_{2}}$ just appears in the vicinity of $\Delta_{1}=\Delta_{2}\simeq 0$.
It is interesting that, in certain regions, the stationary near entanglements $E_{a_{1}a_{2}}$ and $E_{a_{1}m}$ vanish (emerge), while the distant entanglement $E_{a_{2}m}$ emerges (vanishes).
That can provide the platform for the perfect exchange between the near and distant entanglements of enhancement by adjusting the relevant detunings.
To show the entanglement exchange between the three subsystems, we plot $E_{a_{1}a_{2}}$, $E_{a_{1}m}$ and $E_{a_{2}m}$ against $\Delta_{1}$ with $\Delta_{m}=0.87\omega_{b}$ and $0.06\omega_{b}$ in Figs.~\ref{fig6}(d) and~\ref{fig6}(e), respectively.
In Fig.~\ref{fig6}(d), the three entanglements are different, and especially, only the distant entanglement $E_{a_{2}m}$ exists with the suppressed near entanglements $E_{a_{1}a_{2}}$ and $E_{a_{1}m}$ in the range $\Delta_{1}/\kappa_{1}\in[-0.907,-0.943]$; however, figure~\ref{fig6}(e) shows only the nonzero near entanglement $E_{a_{1}m}$. That implies the perfect exchange from the distant entanglement $E_{a_{2}m}$ to the near one $E_{a_{1}m}$ just by adjusting $\Delta_{m}$ from $0.87\omega_{b}$ to $0.06\omega_{b}$.

\begin{figure}[tb]
\centering
\includegraphics[width=8.4 cm]{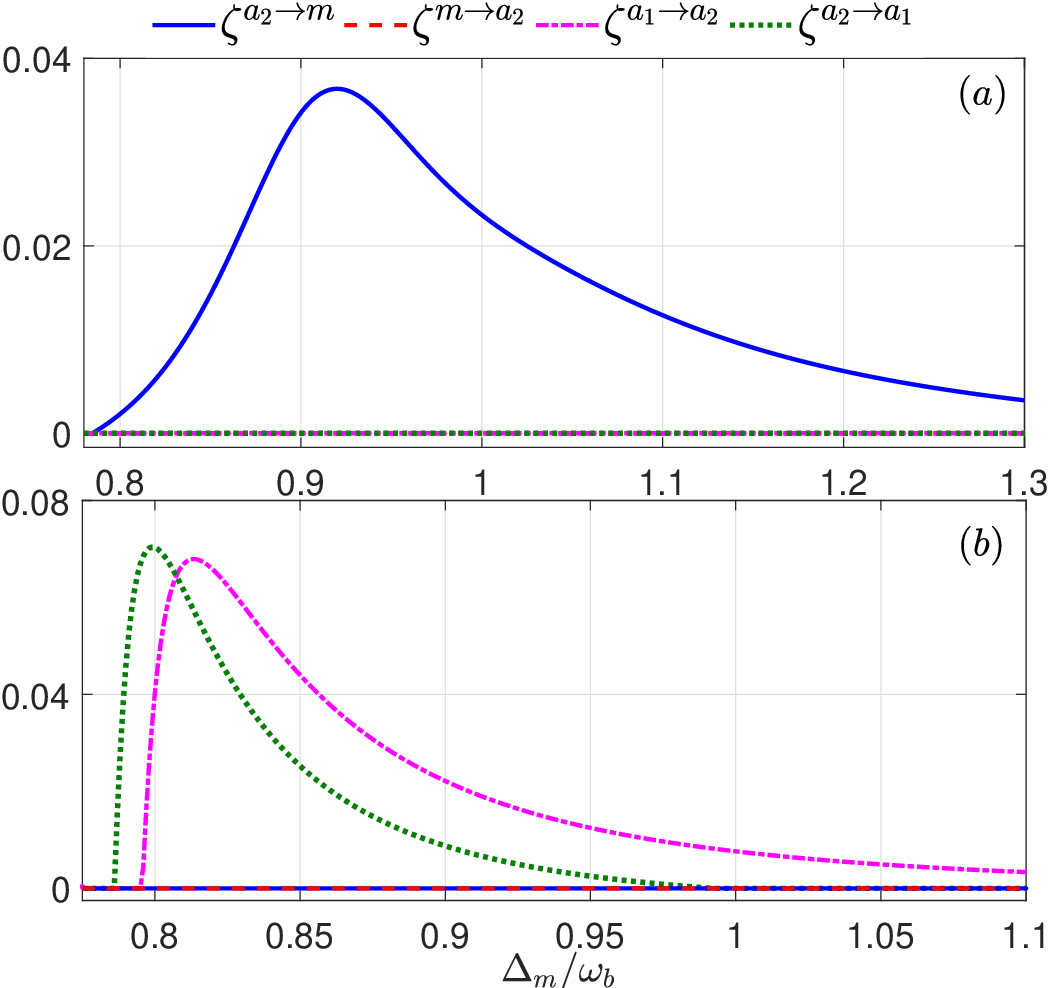}
\caption{(Color online) The quantum steering $\zeta^{a_{2}\rightarrow m}$, $\zeta^{m\rightarrow a_{2}}$, $\zeta^{a_{1}\rightarrow a_{2}}$ and $\zeta^{a_{2}\rightarrow a_{1}}$ versus $\Delta_{m}$ with (a) $\Delta_{1}=-0.96\omega_{b}$ and (b) $\Delta_{1}=-0.13\omega_{b}$. The other parameters are the same as those in Fig.~\ref{fig2}.}
\label{fig7}
\end{figure}

The adjustment of detunings also can modulate the asymmetric EPR steerings.
Figure~\ref{fig7} provides the steerings as the function of detuning $\Delta_{m}$ in subsystems, in which the steering between $a_{1}$ and $m$ is too tiny to be shown here.
As shown in Fig.~\ref{fig7}(a), with $\Delta_{1}=-0.96\omega_{b}$ , there is a distant directional steering $a_{2} \rightarrow m$ without the steering between two photon modes $a_{1}$ and $a_{2}$ in the range $\Delta_{m}/\omega_{b}\in [0.8, 1.3]$.
As shown in Fig.~\ref{fig7}(b), by tuning $\Delta_{1}$ to $-0.13\omega_{b}$, the steering between $a_{2}$ and $m$ disappears, but the asymmetric two-way steerability between $a_{1}$ and $a_{2}$ appears. That means we can realize the exchange between the distant directional steering in the $a_{2}-m$ subsystem and the near asymmetric two-way (even directional) steering in the $a_{1}-a_{2}$ subsystem just by adjusting the detuning.
Thus, this tunable scheme with transfer/exchange of entanglements/steerings might be used to engineer the wanted entanglement/steering between optional modes in the steady state by adjusting relevant parameters.

As mentioned above, the introduction of an auxiliary cavity is critical for generating the distant entanglement/steering and transferring them. Based on the entanglement/steering transfer (exchange) behaviours,
as a basic component, this system can be extended to a cavity lattice/chain involving magnons and phonons via coupling more auxiliary cavities.
In such a chain, a more distant entanglement/steering between the YIG sphere in one side and a further cavity may appear.
That is, the entanglement/steering can transfer through this communicating chain to a greater distance, which is beneficial to the distant transfer of quantum information and remote quantum modulation in the quantum networking \cite{genes2008robust} with a relatively high efficiency.
In addition, recently, we notice a new work \cite{tan2020einstein} that also realizes the EPR entanglement and steering in a mechanical-magnonic cavity system.
To a certain extent, it can demonstrate the reliability of our scheme for the entanglement/steering.

Finally, as for detecting the generated entanglement in the experiment, we  introduce one feasible way to measure the steady-state entanglement via some observable quantities, such as the cavity output modes.
Firstly, according to Eq.~(\ref{eq9}), the entanglement could be deduced based on the CM $V$, every entry of which can be measured from the cavity output mode spectrum.
The entries of the CM of bosonic modes are real and can be precisely extracted via the measurement of the orthogonal quadratures of light mode, which has been realized in the experiment \cite{2005JOB}.
Generally, the arbitrary quadratures of light mode can be obtained by the homodyne detection of the cavity output using a local oscillator with an appropriate phase.
In our system, there are three types of modes as the cavity photon, magnon and mechanical phonon.
The photonic quadratures could be read out directly by homodyning the cavity output.
The quadratures of magnon can be measured by homodyning an additional cavity coupled to the magnon and driven by a weak microwave probe field.
For the magnon mode, an additional cavity of operator $c_{1}$ and frequency $\omega_{c_{1}}$ is used to couple the magnon that has been driven by magnetic field $\omega_{0}$, and then the magnon quadratures can be measured by homodyning the cavity $c_{1}$ probed by a weak microwave field. The dynamics is given as
\begin{equation}
\begin{split}
\dot{c_{1}}=-[i(\omega_{c_{1}}-\omega_{0})+\kappa_{c_{1}}]c_{1}-ig_{mc}m+\sqrt{2\kappa_{c_{1}}}c_{1in},
\label{eq11}
\end{split}
\end{equation}
where $\kappa_{c_{1}}$, $g_{mc}$, $c_{1in}$ are, respectively, the relevant decay rate, coupling strength, and input noise.
Using input-output relation $c_{1out}=\sqrt{2\kappa_{c_{1}}}c_{1}-c_{1in}$ \cite{1985input}, we get
\begin{equation}
\begin{split}
c_{1out}=-(i\sqrt{2}g_{mc}/\sqrt{\kappa_{c_{1}}})m+c_{1in}.
\label{eq12}
\end{split}
\end{equation}
Then, the beam-splitter interaction causes a state swap between the magnon and cavity modes, which can map the magnon quadratures into the cavity output $c_{1out}$.
Thus, we can get the magnon quadratures via measuring $c_{1out}$.
The process of measuring the mechanical quadratures is similar to the magnon case. Via coupling the YIG sphere with an additional cavity $c_{2}$ of frequency $\omega_{c_{2}}$.
A red-detuned field of frequency $\omega_{d}$, i.e., $\omega_{c_{2}}-\omega_{d}=\omega_{b}$, drives the cavity to induce the beam-splitter interaction but suppress the other interaction.
Then the consequent state swap maps the mechanical quadratures into the cavity output.
The relevant analytical expressions are
\begin{equation}
\begin{split}
\dot{c_{2}}=-(i\omega_{b}+\kappa_{c_{2}})c_{2}-ig_{bc}c_{2}(b^{\dagger}+b)+\sqrt{2\kappa_{c_{2}}}c_{2in},
\label{eq13}
\end{split}
\end{equation}
where $\kappa_{c_{2}}$, $g_{bc}$, $c_{2in}$ are, respectively, the relevant decay rate, coupling strength, and the input noise.
 Using input-output relation, we have
\begin{equation}
\begin{split}
c_{2out}=-(i\sqrt{2}g_{bc}|c_2|/\sqrt{\kappa_{c_{2}}})b+c_{2in}.
\label{eq14}
\end{split}
\end{equation}
When $c_{2}$ is driven by a much weaker intra-cavity field, its backaction on the mechanical mode can be neglected. Thus, with Eq. (6), $c_{2out}$ can show the mechanical dynamics.
Then, via measuring the correlations between above cavity outputs, one can extract all entries of CM for yielding the logarithmic negativity as a measurement of the entanglement \cite{2007entanglement,2013mesoscopic}.
In addition, there is another type of entanglement called output entanglement describing the optomechanical entanglement between the experimentally detectable output fields of the cavity and the vibrator in optomechanical systems \cite{genes2008robust,wang2015bipartite,yan2017enhanced}.
Such a straightforwardly detectable entanglement in the magnomechanics is worth discussing in future work.

\section{Conclusions}

In the stable parameter regime of a passive-active double-cavity magnomechanical system, we have analyzed and modulated the stationary continuous variable entanglement and asymmetric EPR steering between the two photon modes of cavities, the magnon mode and mechanical phonon mode of a YIG sphere. In such a well-designed system, the magnon is directly coupled with only the passive cavity that is also coupled with an active cavity via the tunneling effect for the distant entanglement and steering.
Results show the natural magnetostrictive magnon-phonon interaction within the YIG sphere, i.e., the two-mode-squeezing interaction, leads to the initial magnon-phonon entanglement based on the magnomechanics.
Then, in the $\mathcal{PT}$-symmetric-like scheme, the incoherent gain of the auxiliary active cavity can strengthen the effective nonlinear magnomechanical coupling so as to obviously enhance entanglements and improve their robustness against the environment temperature.
And the special features of the $\mathcal{PT}$-symmetric-like structure can satisfy the requirements of the creation of the relatively obvious asymmetric two-way (even directional) distant steering.
In such a tunable system, with the adjustment of the coupling strengths, we can realize the perfect transfer between near and distant entanglements (directional steerings) of different-type modes; both of the entanglements and steerings can also exchange between different subsystems by adjusting detunings.
Thus, the entanglements/steerings can be dynamically switched between different two-mode pairs.
Our work presents a feasible method for genarating the enhanced distant entanglement and EPR steering involving magnons and realizing their transfer/exchange, which may be used in the quantum networks and information processing.
Furthermore, it can be found that, with such a system, the process involving the entanglement/steering generation, transfer, and storage can be designed; in that process, the mechanical oscillator with a low decay relative to the magnon and cavity can show the storage of information.
Those results may pave the way to study entanglements and steerings on the macroscopic scale and more intriguing quantum phenomena based on the magnomechanics, providing an ideal playground for feasible multiple quantum modulations.

\section*{Acknowledgments}

We thanks Prof. Yong Li, Prof. Qiongyi He, Prof. Xiaobo Yan and Dr. Shasha Zheng for helpful discussions. This work is supported by National Natural Science Foundation of China (No. 11704064 and No. 12074061), Fundamental Research Funds for the Central Universities (No. 2412019FZ045) and Science Foundation of the Education Department of Jilin Province (JJKH20211279KJ).


\section*{APPENDIX: THE STEADY-STATE ENTANGLEMENT AND STEERING}
\appendix
\setcounter{equation}{0}
\renewcommand{\theequation}{A\arabic{equation}}
By virtue of the driving field $B_{0}$ and the magnon-photon beam-splitter interaction, we have large cavity and magnon mode amplitudes $|\langle a_{i}\rangle, \langle m\rangle|\gg1$.
Therefore, we can substitute the operators with their mean values plus small fluctuations in Eq.~(\ref{eq2}), i.e., $\aleph(t)=\langle \aleph \rangle+\delta \aleph(t)$ in which $\aleph\equiv(a_{i},m,q,p)$.
Then, the steady-state value of the operators are obtained as
\begin{equation}
\begin{split}
&\langle m\rangle=\frac{\varepsilon_{d}(J^{2}+f_{1}f_{2})}{J^{2}f_{m}+f_{1}f_{2}f_{m}+g_{ma}^{2}f_{2}}\\
&\langle a_{1}\rangle=-\frac{ig_{ma}f_{1}f_{2}\langle m\rangle}{(J^{2}+f_{1}f_{2})f_{1}}\\
& \langle a_{2}\rangle=-\frac{iJ\langle a_{1}\rangle}{f_{2}}\\
&\langle p\rangle=0\\
&\langle q\rangle=-\frac{g_{mb}}{\omega_{b}}|\langle m\rangle|^{2},
\label{eq4}
\end{split}
\end{equation}
in which  $f_{1}=i\Delta_{1}+\kappa_{1}$, $f_{2}=i\Delta_{2}+\kappa_{2t}$, $f_{m}=i\Delta_{m}+\kappa_{m}$, and $\Delta_{m}=\Delta_{m}'+g_{mb}\langle q\rangle$ represents the effective detuning of the magnon mode.
Note that we aim at the dynamic of the quantum fluctuations of the system.
By introducing the orthogonal components as
 \begin{equation}
\begin{split}
 &\delta I_{i}=(\delta a_{i}+\delta a_{i}^{\dagger})/\sqrt{2},~ \delta \varphi_{i}=i(\delta a_{i}^{\dagger}-\delta a_{i})/\sqrt{2},\\
 &\delta x=(\delta m+\delta m^{\dagger})/\sqrt{2},~ \delta y=i(\delta m^{\dagger}-\delta m)/\sqrt{2},\\
 &\delta q=(\delta b+\delta b^{\dagger})/\sqrt{2},~ \delta p=i(\delta b^{\dagger}-\delta b)/\sqrt{2},
\label{eq5}
\end{split}
\end{equation}
we define the vector of quadratures as $W(t)=[\delta I_{1}(t),\delta \varphi_{1}(t),\delta I_{2}(t), \delta \varphi_{2}(t), \delta x(t), \delta y(t), \delta q(t), \delta p(t)]^{T}$ and the noise vectors as $\mu(t)=[\sqrt{2\kappa_{1}}I_{1}^{in}(t), \sqrt{2\kappa_{1}}\varphi_{1}^{in}(t), \sqrt{2\kappa_{2t}}I_{2}^{in}(t), \sqrt{2\kappa_{2t}}\varphi_{2}^{in}(t), \sqrt{2\kappa_{m}}\\x^{in}(t), \sqrt{2\kappa_{m}}y^{in}(t), 0, \xi(t)]^{T}$. We obtain the quadratures in the compact form
\begin{equation}
\begin{split}
\dot{W}(t)=AW(t)+\mu(t),
\label{eq6}
\end{split}
\end{equation}
corresponding to
\begin{small}
\begin{equation}
A=\begin{pmatrix}
-\kappa_{1} & \Delta_{1} & 0 & J & 0 & g_{ma} & 0 &0\\
-\Delta_{1} & -\kappa_{1} & -J & 0 & -g_{ma} & 0 & 0 & 0\\
0 & J & -\kappa_{2t} & \Delta_{2} & 0 & 0 & 0 &0\\
-J & 0 & -\Delta_{2} & -\kappa_{2t} & 0 & 0 & 0 & 0\\
0 & g_{ma} & 0 & 0 & -\kappa_{m} &\Delta_{m} & -G_{mb} & 0\\
-g_{ma} & 0 & 0 & 0 &\Delta_{m} & -\kappa_{m} & 0 & 0\\
0 & 0 & 0 & 0 & 0 & 0 & 0 & \omega_{b}\\
0 & 0 & 0 & 0 & 0 & G_{mb} & -\omega_{b} & -\gamma_{b}\
\label{eq7}
\end{pmatrix},
\end{equation}
\end{small}where $G_{mb}=i\sqrt{2}g_{mb}\langle m\rangle$ is the effective magnomechanical coupling between the magnon and phonon modes. Generally, the covariance matrix (CM) of $n$ bosonic modes is a real and symmetric matrix in $2n$ dimension \cite{giedke2002characterization,pirandola2009correlation,kogias2015quantification}.
In our system, there are four modes (two photon modes, a mangon mode and a phonon mode), which can be written as eight quadrature operators as shown in Eq.~(\ref{eq5}), satisfying the specific commutation relation. Thus, as a continuous variable four-mode Gaussian state, the system can be completely characterized by a $8\times8$ CM $V$.
The steady-state $V$ can be straightforwardly obtained by solving the Lyapunov equation \cite{2007entanglement}
\begin{equation}
\begin{split}
AV+VA^{T}=-D.
\label{eq8}
\end{split}
\end{equation}

For steady-state entanglements and steerings, the steady-state CM $V$ is required to be asymptotically stable, and thus the eigenvalues of the Jacobian matrix $A$ as Eq.~(\ref{eq7}) need to be solved first. With a set of parameters, the system is stable only if the real parts of all eigenvalues are negative \cite{dumeige2011stability}; and then we use CM $V$ from Eq.~(\ref{eq8}) to yield the quantities for quantifying the steady-state entanglement within the stable parameter regime. Otherwise, the system is unstable and then the relevant parameters are not used.

We define $V$ as $V_{ij}=\langle v_{i}(t)v_{j}(t')+v_{j}(t')v_{i}(t)\rangle/2$ $(i,j=1,2,..., 8)$ and $D$=diag$[\kappa_{1}(2N_{a_{1}}+1),\kappa_{1}(2N_{a_{1}}+1),\kappa_{2t}(2N_{a_{2}}+1),\kappa_{2t}(2N_{a_{2}}+1),\kappa_{m}(2N_{m}+1),\kappa_{m}(2N_{m}+1),0,\gamma_{b}(2N_{b}+1)]$, in which the diffusion matrix $D$ are defined by $\delta(t-t')D_{i,j}=\frac{1}{2}\langle n_{i}(t)n_{j}^{\dagger}(t')+n_{j}^{\dagger}(t)n_{i}(t')\rangle$.
Equation~(\ref{eq8}) can be straightforwardly solved, however, it is too complicated to give the exact expression. Thus, under local operations, classical communications and an upper bound for the distillable entanglement \cite{2002}, we consider the logarithmic negativity $E_{M_{1}M_{2}}$ to quantify the entanglement between mode $M_1$ and mode $M_2$ as \cite{2004}
\begin{equation}
\begin{split}
E_{M_{1}M_{2}}\equiv max[0, -ln2\tilde{v}_{-}].
\label{eq9}
\end{split}
\end{equation}
Here $\tilde{v}_{-}$=min[eig($i\Omega P \mathcal{V}_{4}P$)], where $\Omega=\oplus_{j=1}^{2}i\sigma_{y}$ with the Pauli matrix $\sigma_{y}$ , $\mathcal{V}_{4}$ is the $4\times4$ CM of the under-researched modes $M_1$ and $M_2$ by removing the unwanted rows and columns of other modes in $V$, and $P$=diag(1, -1, 1, 1) is the matrix that performs partial transposition on CM \cite{2000}.

For the discussion of the EPR steering, the proposed measurement of the Gaussian steerability in different directions between mode $M_1$ and mode $M_2$ are \cite{steering3,steering5}
\begin{equation}
\begin{split}
&\mathcal{V}_4=\begin{pmatrix}
A&B\\
B^{T}&C\
\end{pmatrix},\\
&\zeta^{M_{1}\rightarrow M_{2}}=max\{0, S(2A)-S(2\mathcal{V}_4)\},\\
&\zeta^{M_{2}\rightarrow M_{1}}=max\{0, S(2C)-S(2\mathcal{V}_4)\},\\
\label{eq10}
\end{split}
\end{equation}
with $S(\rho)=\frac{1}{2}$lndet($\rho$).


\end{document}